\documentclass[%
 reprint,
 superscriptaddress,
 amsmath,amssymb,
 aps,prc,
]{revtex4-2}

\usepackage{graphicx}

\usepackage{bm}
\usepackage{booktabs}

\begin{document}

\title{$\alpha$--$\alpha$ Correlations in Excited Clustering Nuclei Across the Solid-Liquid Transition
}

\author{Yong-Hao Jin}
\affiliation{Shanghai Institute of Applied Physics, Chinese Academy of Sciences, Shanghai 201800, China}
\affiliation{Shanghai Advanced Research Institute, Chinese Academy of Sciences, Shanghai 201210, China}
\affiliation{University of Chinese Academy of Sciences, Beijing 101408, China}

\author{Xi-Guang Cao}
\email[Contact author: ]{caoxg@sari.ac.cn}
\affiliation{Shanghai Advanced Research Institute, Chinese Academy of Sciences, Shanghai 201210, China}
\affiliation{Shanghai Institute of Applied Physics, Chinese Academy of Sciences, Shanghai 201800, China}
\affiliation{University of Chinese Academy of Sciences, Beijing 101408, China}

\author{Yu-Gang Ma}
\email[Contact author: ]{mayugang@fudan.edu.cn}
\affiliation{Key Laboratory of Nuclear Physics and Ion-beam Application (MOE), Institute of Modern Physics, Fudan University, Shanghai 200433, China}
\affiliation{Shanghai Research Center for Theoretical Nuclear Physics, NSFC and Fudan University, Shanghai 200438, China}
\affiliation{School of Physics, East China Normal University, Shanghai 200241, China}

\begin{abstract}
We investigate the structural evolution of $\alpha$-cluster matter using a microscopic transport model for four $\alpha$-conjugate collision systems, including $^{12}\text{C}+^{12}\text{C}$, $^{16}\text{O}+^{12}\text{C}$, $^{28}\text{Si}+^{12}\text{C}$, and $^{40}\text{Ca}+^{12}\text{C}$ at incident energies ranging from 2.0 to 7.0~MeV/u.
The $\alpha$-cluster structure of an excited nucleus undergoes a solid-liquid phase transition with increasing incident energy.
Whether it proceeds by the dissociation of individual clusters or by the collapse of their spatial organization remains unexplored, as does the question of whether any characteristic $\alpha$--$\alpha$ correlation survives.
The one-body density $\rho(\mathbf{r})$, the standard clustering diagnostic, cannot by itself settle this question, since it neither distinguishes a genuine cluster configuration from a density fluctuation nor reveals the sparse residual clusters masked by the surrounding nucleon medium.
We therefore construct the exclusive $\alpha$--$\alpha$ two-body spatial correlation function from the $\alpha$-cluster centroids to isolate the $\alpha$ clusters.
In all four systems the $\alpha$--$\alpha$ correlation exhibits two peaks, a nearest-neighbor one and a next-nearest-neighbor inter-cluster one. Both peak positions increase with system size.
Over the energy range studied, which spans this transition, the mean number of $\alpha$ clusters per event falls steeply while the $\alpha$--$\alpha$ correlation persists and the inter-cluster distances change only slightly.
The individual clusters therefore dissociate rather than lose their spatial organization.
\end{abstract}

\maketitle


\section{Introduction}\label{sec:introduction}

The cluster picture of the nucleus was established in the early history of nuclear physics in the 1930s~\cite{PhysRev.54.681, Bethe1936}.
It is supported by the inertness of the $\alpha$ particle, the $\alpha$ decay, and the linear relation between binding energies of $\alpha$ conjugate nuclei and the hypothetical $\alpha-\alpha$ bond numbers.
Ab initio calculations show that $\alpha$ clustering occurs in the ground and excited states of light nuclei, such as $^{8,10}$Be and $^{12}$C~\cite{otsuka2022alpha}.
The development of new measurement methods with unprecedented accuracy is also driven by open questions in cluster physics, ranging from exotic cluster configurations~\cite{He:2014iqa, Cao:2018cvf, Xu2025PRC, Xu2026PRC} to the possible nucleon-level Bose-Einstein condensation of $\alpha$ clusters~\cite{ZhouB, PRL07} and to the role of cluster structures in the equation of state of low-density astrophysical matter~\cite{PRL12}. Dedicated charged-particle telescope arrays have been developed to pursue them~\cite{Liao2024NST, xu_charge2026}.
Previous investigations have predominantly focused on the static cluster configurations of ground states or specific low-lying excited states, including simple $\alpha$-$\alpha$ structures, binary $\alpha$+core molecular-like structures, and near-threshold resonant states such as the Hoyle state~\cite{Freer_2007, vonoertzen2006, FreerFynbo2014}.

Experimentally, $\alpha$-clustering in light nuclei is well established.
The Hoyle state in $^{12}$C and the excited states of $^{16}$O exhibit pronounced triangular and tetrahedral $\alpha$-cluster configurations, respectively~\cite{PhysRevLett.113.012502, Bijker:2014tka, PhysRevLett.112.102501}.
These localized structures have been verified through various experimental signatures, including enhanced electric monopole ($E0$) and quadrupole ($E2$) transition strengths, $\alpha$-knockout and transfer reactions, and the observation of rotational bands characterized by exceptionally large moments of inertia~\cite{Freer_2007}.

A further question is whether $\alpha$-cluster geometric ordering can persist beyond the light-mass region.
Medium-mass $N=Z$ nuclei, with their larger number of constituent $\alpha$ particles, offer the possibility of more complex multi-cluster configurations.
In $^{28}$Si, high-excitation states that decay into $7\alpha$ have been identified~\cite{Natowitz2026PRC}, indicating that the entire $\alpha$ content of a medium-mass nucleus can be organized into a single cluster configuration.

Among these, the $10\alpha$ conjugate nucleus $^{40}\text{Ca}$ is of particular interest, because microscopic calculations find $\alpha$--$^{36}$Ar and $^{12}$C--$^{28}$Si cluster configurations in its excited states~\cite{KanadaEnyo2005PRC,Taniguchi2007PRC}, and its $\alpha$-cluster structures are well established~\cite{Ren2018FrontPhys}.
However, the conventional shell model has historically described this benchmark doubly magic nucleus as a rigid, spherical, and homogeneous independent-particle system.
From a cluster-model perspective, it can likewise be envisioned as a complex compound system of ten $\alpha$ particles.
Over the decades, evidence from rotational bands, highly deformed states, and $\alpha$-transfer reactions has suggested the existence of cluster degrees of freedom in the excited states of $^{40}\text{Ca}$~\cite{YAMAYA1994154}.
Yet the vast majority of these investigations have been confined to static properties of low-lying states or extreme asymptotic limits.

The cluster structure undergoes a dramatic transition as the incident energy increases.
Whether this transition involves the dissolution of individual clusters, the loss of their spatial organization, or both remains to be established.
Highly excited cluster states are therefore studied to characterize this structural change and, in particular, to determine what remains of the cluster structure after long-range correlations between clusters are lost.

Accordingly, the two-body spatial correlation function $P(r)$ is defined as the statistical distribution of relative distances between any two constituents (Sec.~\ref{sec:Pr}).
It can distinguish localized aggregation, molecular-like alignments, and more diffuse nonlocal correlations.
The two-body correlation is inherently better suited than the one-body density distribution to revealing non-trivial correlations among clusters.
It is therefore an ideal tool for probing the structural change from molecular-like configurations to a cluster-depleted state.
This structural transition is a non-equilibrium dynamical process driven by the competition between attractive nuclear forces and thermal fluctuations.
As the incident energy increases (e.g., within the range of 2.0--7.0~MeV/u), the enhanced thermal motion and collision-induced fluctuations are expected to disrupt the geometric order and drive the nucleons out of their cluster-localized configurations.

Describing this process requires a framework that can handle both localized cluster formations and the independent-particle continuum without a priori structural assumptions.
Traditional static models, such as constrained Hartree-Fock calculations~\cite{Ebran_2014PRC}, provide a detailed picture of the equilibrium cluster configurations but are less suited for tracking how these configurations evolve with incident energy.
The Extended Quantum Molecular Dynamics (EQMD) model addresses this limitation by explicitly treating each nucleon as a spatially localized wave packet.
It naturally incorporates clusterization and dynamical transport.

In our previous work~\cite{SLPT2026}, we used EQMD calculations to establish a generic solid-liquid phase transition in $\alpha$-cluster matter across the four $\alpha$-conjugate collision systems $^{12}\text{C}+^{12}\text{C}$, $^{16}\text{O}+^{12}\text{C}$, $^{28}\text{Si}+^{12}\text{C}$, and $^{40}\text{Ca}+^{12}\text{C}$.
This conclusion was based on thermodynamic and informational observables, and the inclusive two-body correlation function $P(r)$ provided a direct spatial signature of the transition.
The present work extends this line by introducing the exclusive $\alpha$--$\alpha$ correlation channel, which isolates pairs of $\alpha$ clusters from pairs of all other kinds.
This allows us to establish the physical origin of each correlation peak, to track changes in the spatial order among the $\alpha$ clusters across the transition, and to see how that order depends on the size of the system.

\section{Methods}

We simulate central collisions of $\alpha$-conjugate nuclei with the EQMD model.
The four systems studied are $^{12}\text{C}+^{12}\text{C}$, $^{16}\text{O}+^{12}\text{C}$, $^{28}\text{Si}+^{12}\text{C}$, and $^{40}\text{Ca}+^{12}\text{C}$.
The simulations use incident energies ranging from 2.0 to 7.0~MeV/u and impact parameters $b \leq 1$~fm. For $^{40}$Ca+$^{12}$C the analysis starts at 2.5~MeV/u.
In the following, we describe the EQMD framework employed in this work.

\subsection{EQMD transport model}
\label{sec:EQMD}

The EQMD model~\cite{Maruyama:1995dc} was formed on the basis of the Quantum Molecular Dynamics (QMD) model by improving the description for the nuclear ground state, which considerably influences heavy-ion reactions at low and intermediate energies. Three modifications distinguish it from the standard QMD approach.
First, a Pauli potential is introduced to approximate the fermionic nature of the nuclear many-body system.
It prevents nucleons of the same spin and isospin from approaching each other arbitrarily closely, as they would otherwise be free to do in QMD.
Second, the width of each nucleon is not a constant shared by all nucleons but an independent dynamical variable, whose real and imaginary parts correspond to the widths in coordinate space and momentum space, respectively, and whose evolution follows from the time-dependent variational principle.
Third, a single parameter set can reproduce nuclear ground-state properties well over a wide mass range~\cite{wangssNPR_2015}, and the ground state is obtained by frictional cooling, which enforces the Pauli principle and renders the initialized nuclei stable against spurious emission of nucleons or fragments over arbitrarily long evolution times.

The quasi-classical EQMD model does not include Bose statistics. The $\alpha$ clusters analyzed here are therefore treated as classical configurations of nucleons and identified using the procedure described in Sec.~\ref{sec:cluster}.

The EQMD model has been successfully applied to a series of reactions with different reaction mechanisms spanning wide energy regions, such as the giant resonance of nuclei with exotic cluster configurations~\cite{He:2014iqa,Huang:2021mct}, large deformations~\cite{Wang:2017tfr}, quasi-deuteron reaction induced by high energy $\gamma$ ray~\cite{HBS95PRC}, and clusters behaviors in relativistic heavy-ion collisions~\cite{ZhangS95PRC,MaYG:2022dbh}, etc. The same model has also been applied to bubble structures in light nuclei, including their breathing modes~\cite{REN2024PLB}, and to a unified classification and quantification of bubble-like nuclei~\cite{REN_2026NST}.
These applications show that the $\alpha$-cluster configurations produced by EQMD remain consistent with experimental observations across a range of reaction mechanisms~\cite{He:2014iqa,He:2016cwt,Dey_PRC_2020,Huang:2021mct}, which underpins its use here to follow the spatial organization of $\alpha$ clusters in an excited nucleus.

Within this framework, the total wave function is expressed as a direct product of single-nucleon wave functions.

\begin{equation}
    \Psi=\prod_{i}\phi_{i}(\mathbf{r}_{i}),
\end{equation}

\begin{equation}
    \phi_{i}(\mathbf{r}_{i})=\left(\frac{\upsilon_{i}+\upsilon^\ast_{i}}{2\pi}\right)^{\frac{3}{4}}\exp\left[-\frac{\upsilon_{i}}{2}
    (\mathbf{r}_{i}-\mathbf{R}_{i})^2+\frac{i}{\hbar}\mathbf{P}_{i}\cdot \mathbf{r}_{i}\right],
    \label{eq:wavefunction_single}
\end{equation}
in which $\mathbf{R}_{i}$ and $\mathbf{P}_{i}$ are the centroids of the wave packet belonging to the $i$th nucleon. The Gaussian width $\upsilon_{i}$ is introduced as a complex quantity,
\begin{equation}
    \upsilon_{i}=\frac{1}{\lambda_{i}}+i\delta_{i}.
\end{equation}

The one-body nucleon density distribution $\rho_i(\mathbf{r})$ is obtained from the Wigner transform of the single-particle wave function. The Wigner transform of Eq.~(\ref{eq:wavefunction_single}) yields the phase-space distribution
\begin{equation}
    \begin{aligned}
    f(\mathbf{r}, \mathbf{p}) &= \frac{1}{(2\pi\hbar)^3}
        \int \exp\left( \frac{i\mathbf{p} \cdot \bm{\xi}}{\hbar} \right)
        \phi_i\!\left(\mathbf{r} - \frac{\bm{\xi}}{2}\right)
        \phi_i^*\!\left(\mathbf{r} + \frac{\bm{\xi}}{2}\right) \,\mathrm{d}\bm{\xi} \\[4pt]
    &= \frac{1}{(\pi\hbar)^3}
        \exp\left[ -\frac{1 + \lambda_i^2\delta_i^2}{\lambda_i}
        (\mathbf{r} - \mathbf{R}_i)^2 \right] \\
    &\quad \times \exp\left[ -\frac{\lambda_i}{\hbar^2}
        (\mathbf{p} - \mathbf{P}_i)^2 \right]
        \exp\left[ \frac{2\lambda_i\delta_i}{\hbar}
        (\mathbf{r} - \mathbf{R}_i)(\mathbf{p} - \mathbf{P}_i) \right].
    \end{aligned}
    \label{eq:wigner_transform}
\end{equation}
Integrating over momentum space gives the single-particle spatial density
\begin{equation}
    \rho_i(\mathbf{r}) = \int f(\mathbf{r}, \mathbf{p})\, d\mathbf{p}
    = \left( \frac{1}{\pi\lambda_i} \right)^{3/2}
    \exp\left[ -\frac{(\mathbf{r} - \mathbf{R}_i)^2}{\lambda_i} \right],
    \label{eq:single_particle_density}
\end{equation}
and the total density of the system is the superposition of all single-nucleon densities, $\rho(\mathbf{r}) = \sum_i \rho_i(\mathbf{r})$. This expression provides the fundamental link between the microscopic EQMD wave-packet dynamics and the macroscopic density distributions discussed in Sec.~\ref{sec:density}.

The system's Hamiltonian is written as
\begin{equation}
    \begin{split}
        H&=\langle\psi\vert\sum_i-\frac{\hbar^2}{2m}\nabla_{i}^2-\hat{T}_{\mathrm{c.m.}}
        +\hat{H}_{\mathrm{int}}\vert\psi\rangle\\
        &=\sum_{i}\left[\frac{\mathbf{P}_{i}^2}{2m}+\frac{3\hbar^2
        (1+\lambda_{i}^2\delta_{i}^2)}{4m\lambda_{i}}\right]-T_{\mathrm{c.m.}}+H_{\mathrm{int}},
    \end{split}
\end{equation}
in which a zero-point center-of-mass kinetic energy $T_{\mathrm{c.m.}}$ is subtracted because it is not a constant value, receiving different contributions from the dynamical wave-packet widths compared with the traditional QMD model.
\begin{equation}
    T_{\mathrm{c.m.}}=\sum_{i}\frac{t_{i}^{\mathrm{c.m.}}}{M_{i}},
\end{equation}
where $t_{\mathrm{c.m.}}$ is zero-point center-of-mass kinetic energy contributed by the $i$th nucleon, and $M_{i}$ is an effective mass number of a fragment. They are written as
\begin{equation}
    t_{i}^{\mathrm{c.m.}}=-\frac{\langle\phi_{i}\vert\hbar\nabla^2\vert\phi_{i}
    \rangle}{2m}+\frac{\langle\phi_{i}\vert\hbar\nabla\vert\phi_{i}
    \rangle^2}{2m},
\end{equation}

\begin{equation}
    \begin{split}
    &M_{i}=\sum_{j}F_{ij},\\
    &F_{ij} \equiv \left\{
    \begin{array}{lr}
    1,&\vert\mathbf{R}_{i}-\mathbf{R}_{j}\vert<a\\
    \exp\left[-(\mathbf{R}_{i}-\mathbf{R}_{j})^2/b\right], &\vert\mathbf{R}_{i}-\mathbf{R}_{j}\vert\ge a
    \end{array}
    \right.
    \end{split}
\end{equation}
in which $a=1.7$ fm and $b=4$ fm$^{2}$.

The effective interaction potential of the system can be expressed as follows
\begin{equation}
    H_{\mathrm{int}}=H_{\mathrm{Skyrme}}+H_{\mathrm{Coul}}+H_{\mathrm{Sym}}+H_{\mathrm{Pauli}}.
\end{equation}
The Skyrme interaction can be expressed as
\begin{equation}
    H_{\mathrm{Skyrme}}=\frac{\alpha}{2\rho_{0}}\int\rho^2\,\mathrm{d}^{3} r+\frac{\beta}{(\gamma+1)\rho_{0}^{\gamma}}\int\rho^{\gamma+1}\,\mathrm{d}^{3}r.
\end{equation}
The density is the superposition of all nucleon densities,
\begin{equation}
    \rho(r)=\sum\limits_{i}^{A}\rho_{i}(r),
\end{equation}

\begin{equation}
    \rho(\mathbf{r}_{i})=(\frac{1}{\pi\lambda_{i}})^{\frac{3}{2}}\exp[-\frac{1}{\lambda_{i}}
    (\mathbf{r}_{i}-\mathbf{R}_{i})^2].
\end{equation}
The symmetry potential can be represented as
\begin{equation}
    H_{\mathrm{Sym}}=\frac{c_{S}}{2\rho_{0}}\sum\limits_{i,j\ne i}\int[2\delta(T_{i},T_{j})-1]\rho_{i}(\mathbf{r})\rho_{j}(\mathbf{r})\mathrm{d}^3 r.
\end{equation}
The Pauli potential can be written as
\begin{equation}
    \begin{split}
    H_{\mathrm{Pauli}}=\frac{c_{P}}{2}\sum_{i}(f_{i}-f_{0})^\mu\theta(f_{i}-f_{0}),\\
    f_{i}\equiv\sum_{j}\delta(S_{i},S_{j})\delta(T_{i},T_{j})\vert
    \langle\psi_{i}
    \vert\psi_{j}\rangle\vert,
    \end{split}
\end{equation}
where $f_{i}$ is the overlap of the $i$th nucleon with other nucleons having the same spin and isospin, and $\theta$ is the unit step function. The parameters selected for potential energy are $\alpha=-124.3$ MeV, $\beta=70.5$ MeV, $\gamma=2$, $c_{S}=25$ MeV, $c_{P}=15$ MeV, $f_{0}=1.0$, and $\mu=1.3$.

\begin{table*}[t]
    \caption{The binding energies and root-mean-square (RMS) radii of the initialized nuclei predicted by the EQMD model are listed and compared with experimental values taken from \cite{BE_exp_2020,RMS_exp_2013}.}

    \begin{tabular*}{15cm} {@{\extracolsep{\fill} } ccccc}
    \toprule
    Nuclei & $E/A$ (EQMD) & $E/A$ (EXP) & RMS radius (EQMD) & RMS radius (EXP) \\
     & MeV & MeV & fm & fm  \\
    \midrule
    $^{12}$C  &7.7210  & 7.6801   & 2.313 & 2.470 \\
    $^{16}$O  &7.9265  & 7.9762   & 2.533 & 2.699 \\
    $^{20}$Ne &7.9998  & 8.0322   & 2.783 & 3.005 \\
    $^{24}$Mg &8.0688  & 8.2607   & 2.750 & 3.057 \\
    $^{28}$Si &8.1240  & 8.4477   & 2.853 & 3.122 \\
    $^{32}$S  &8.1473  & 8.4931   & 2.955 & 3.261\\
    $^{40}$Ca &8.1240  & 8.5513   & 3.156 & 3.477 \\
    \bottomrule
    \end{tabular*}
    \label{Table:BE}
\end{table*}

Table~\ref{Table:BE} shows the calculated binding energies and RMS radii of the ground states. The EQMD model reproduces the experimental data fairly satisfactorily, considering that it is a transport model for heavy-ion collisions rather than a dedicated nuclear structure model. The relative fluctuations of the binding energy and RMS radius during free evolution are less than $10^{-2}$ and $10^{-1}$ percent, respectively, which are negligible over the simulation timescale.

The compound system reaches thermal quasi-equilibrium before freeze-out. The calibration of the temperature against incident energy follows our previous study~\cite{SLPT2026}, and the details are not repeated here.

\subsection{Cluster identification and analysis conditions}
\label{sec:cluster}
\label{sec:freezeout}
\label{sec:noemission}

Clusters are identified in coordinate space with a minimum-spanning-tree algorithm, a standard criterion in molecular dynamics calculations~\cite{AICHELIN1991} that underlies the cluster searches of transport models~\cite{mitsyn2015}.
The implementation is that of our previous study~\cite{SLPT2026}.
Nucleons are assigned to a common fragment when their relative distance remains below a fixed linking distance, and each fragment is characterized by its mass number $A$ and charge number $Z$ by the position of its center of mass.
The $\alpha$ clusters analyzed below are those fragments with $A=4$ and $Z=2$, and only events in which the compound system survives intact are retained for the analysis.
The linking criteria are held fixed throughout the analysis, and no systematic sensitivity study of the identification against a change of criteria has been carried out, so that its stability is not established here.

All observables reported below are evaluated at a single freeze-out time, taken as $t=5000$~fm/$c$.
The compound system reaches thermal equilibrium at around several thousand fm/$c$~\cite{SLPT2026}, so that $5000$~fm/$c$ lies in the equilibrated regime, in which the properties of the system no longer vary systematically with the instant of measurement. The same freeze-out time is used for all four systems and for every incident energy.

The collision is evolved with a conserved total Hamiltonian, and no channel for the emission or decay of particles is included in the calculation. The spurious emission of nucleons and fragments is further suppressed by the initialization procedure, which yields nuclei that remain stable over arbitrarily long evolution times~\cite{SLPT2026}. Together with the event selection described above, this defines the sample addressed in this work as complete compound systems, in which the deposited energy is retained within the system throughout the evolution.

\subsection{Two-body spatial correlation function}
\label{sec:Pr}

The spatial structure of this sample is characterized by the two-body spatial correlation function, which is intrinsically more sensitive to multi-body geometric ordering than the one-body density.
Formally, it is defined as the statistical distribution of relative distances $r$ between any two constituents (single nucleons and clusters) of the compound system.

\begin{equation}
    P(r) = \Big\langle \sum_{i<j} \delta(|\mathbf{R}_i - \mathbf{R}_j| - r) \Big\rangle,
    \label{eq:Pr}
\end{equation}
where $\mathbf{R}_i$ and $\mathbf{R}_j$ denote the center-of-mass coordinates of the constituents, and the brackets $\langle \dots \rangle$ indicate the ensemble average over all calculated events.

The function $P(r)$ follows the same construction as the relative two-body density of nuclear many-body theory, $\rho_{\mathrm{rel}}(r)=\int \rho_{2}(\mathbf r,\mathbf R)\,d\mathbf R$, obtained by projecting the two-body density onto the relative coordinate and integrating over the center of mass~\cite{ATTI2015}.
The pairs are taken here over all constituents of the system rather than over nucleons alone, and $P(r)$ differs from that density by the factor $4\pi r^{2}$ and by the normalization.
It is the counterpart of the pair distribution function of liquid-state theory~\cite{HansenMcDonald2013}, from which the radial distribution function $g(r)=P(r)/(4\pi r^{2}\rho)$ is obtained only after a further division by the number density $\rho$ of the constituents.
That division is not applied here, because the mixture of nucleons and clusters does not admit a well-defined number density.
The coordinate-space distance entering Eq.~(\ref{eq:Pr}) is the same one that the fragment-recognition criterion of Sec.~\ref{sec:cluster} employs as a binary cut. Retaining it as a distribution is what preserves the multi-body geometric ordering that the one-body density averages out.
Numerically, $P(r)$ is calculated via the discretized pair summation.

\begin{equation}
    P(r) = \frac{1}{N_{\mathrm{ev}}} \sum_{\mathrm{events}} \frac{1}{M(M-1)} \sum_{i \neq j}^{M} \delta(r - |\mathbf{R}_i - \mathbf{R}_j|),
    \label{eq:Pr_discrete}
\end{equation}
where $M$ denotes the total number of constituents ($M \ge 2$) in a given event, $\mathbf{R}_i$ is their center-of-mass position, and $N_{\mathrm{ev}}$ is the total number of simulated events.
Eq.~(\ref{eq:Pr_discrete}) is the numerical implementation of Eq.~(\ref{eq:Pr}). It averages the per-event pair distribution, normalized by the number of its pairs, over $N_{\mathrm{ev}}$ events, so that events with different $M$ contribute on an equal footing.
Each distribution therefore integrates to unity and has units of inverse length.

In this work, we evaluate two variants of $P(r)$, the \textit{inclusive} $P(r)$ using all constituents, including single nucleons and all clusters, e.g., $d, t, \alpha$, to characterize global structural transitions.
We calculate the \textit{exclusive} $P(r)$ from $\alpha$-cluster centroids alone, so that $\alpha$--$\alpha$ pairs are examined separately.

\section{Results}

\subsection{Density distribution of nuclear matter}
\label{sec:density}
\begin{figure}[htbp]
    \centering
    \includegraphics[width=0.9\columnwidth]{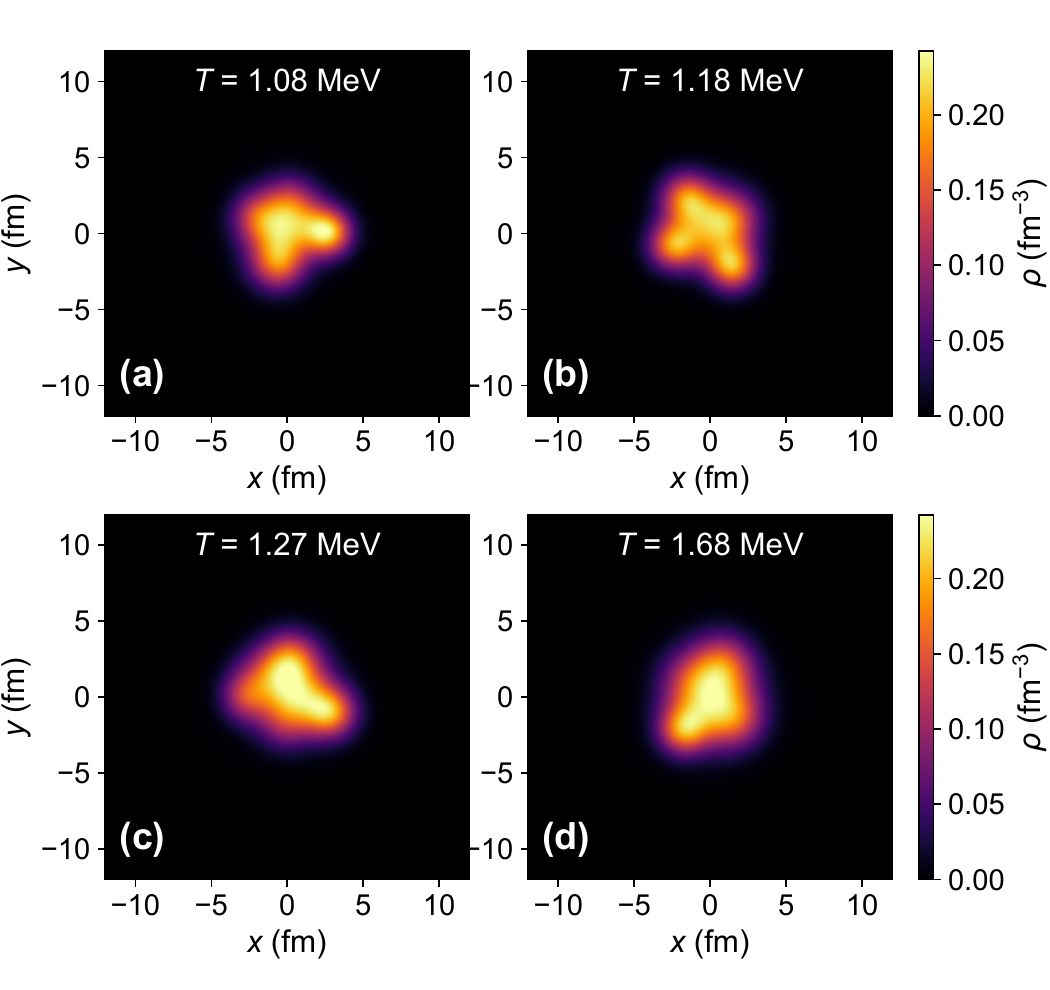}
    \caption{
    Shown is the evolution of the spatial nucleon density distribution of $^{40}\text{Ca}^*$ projected onto the $x$-$y$ plane (at $z=0$) at four representative temperatures. Each panel corresponds to a representative single event.}
    \label{fig:40Carho}
\end{figure}

To characterize the spatial configuration of $^{40}\text{Ca}$ and its evolution with temperature, we first examine the one-body nucleon density distribution.
The same qualitative behavior is found for all four systems, so that $^{40}\text{Ca}^*$ serves as a representative case.
In the EQMD framework, each nucleon is represented as a spatially localized Gaussian wave packet, and the total one-body density $\rho(\mathbf{r})$ is obtained by summing the single-particle densities of all nucleons (see Methods, Sec.~\ref{sec:EQMD} for details).
To visualize the overall spatial arrangement, we project $\rho(\mathbf{r})$ onto the $x$-$y$ plane, transverse to the beam direction.
Figure~\ref{fig:40Carho} displays the density profile of $^{40}\text{Ca}^*$, the compound system formed in $^{28}\text{Si} + ^{12}\text{C}$ collisions, at four representative temperatures spanning the range studied.

The density profiles show no systematic change with temperature.
At each temperature, the nucleons remain confined to a compact region at close to the saturation density of the medium.
Consequently, the profile exhibits only local density maxima separated by regions of moderately lower density.
No extended dilute regions isolate individual clusters, and at no temperature does the profile develop the strongly multi-centered pattern of well-separated cores and low-density necks that a much lower average density would require.
The overall spatial extent of the profile likewise shows no systematic dependence on temperature. The system does not expand appreciably over the range studied (Fig.~\ref{fig:40Carho}). The density is used here only as a qualitative indicator, and no quantitative statement about its behavior is drawn.

The smoothness of $\rho(\mathbf{r})$ is therefore not by itself a reliable criterion for the presence or absence of $\alpha$ clustering, since it would miss the sparse residual clusters masked by the surrounding nucleon medium, and may also report structure arising from local density fluctuations in the absence of any clustering.
Nor can a density near saturation be read as requiring that clusters have dissociated. Short-range-correlated nucleon pairs persist at and above $\rho_0$~\cite{Burrello2022}, and a description that retains only nucleons contains no cluster degrees of freedom by construction.
To establish the existence and evolution of the $\alpha$-cluster spatial order, we therefore turn to the two-body spatial correlation function, which is inherently sensitive to multi-body geometric ordering.

\subsection{Inclusive correlation function}

To place the structural evolution of the present compound system in context, we first recall the inclusive results reported in our previous study~\cite{SLPT2026}.
In our previous study, the inclusive P(r) exhibited a discrete multi-peak structure.
In all four $\alpha$-conjugate systems, the long-range peaks gradually faded and the short-range peak grew as the incident energy increased.
Here, we take this inclusive result as a reference baseline only, since our aim is not to re-examine the melting sequence.

However, the inclusive $P(r)$ alone leaves two questions unresolved. The first concerns the physical origin of each correlation peak in the low-energy solid phase. The second concerns how the $\alpha$ clusters themselves respond to the transition.
To address these questions, we introduce the exclusive $\alpha$--$\alpha$ correlation channel, which retains only pairs of $\alpha$ clusters.
We then analyze the peak structure in detail to assign a characteristic distance to each peak.

\begin{figure}[htbp]
    \centering
    \includegraphics[width=0.95\columnwidth]{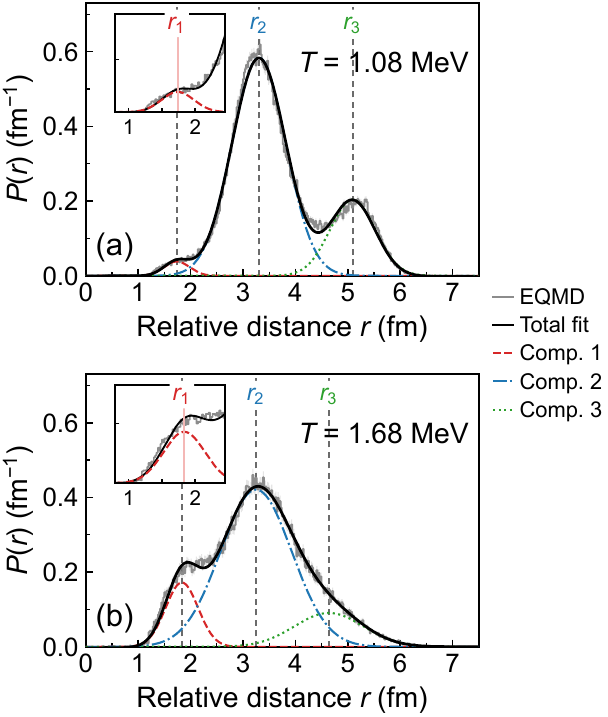}
    \caption{Two-body spatial correlation function $P(r)$ for the $^{28}\text{Si} + ^{12}\text{C}$ system (compound nucleus $^{40}\text{Ca}^*$) at (a) $T = 1.08$~MeV and (b) $T = 1.68$~MeV, corresponding to incident energies of 2.5 and 7.0~MeV/u. The gray stepped lines are the EQMD results, the shaded bands their statistical uncertainties, the black solid lines a three-Gaussian fit plus a linear background, and the colored lines the individual Gaussian components. Vertical dashed lines mark the fitted peak positions. The insets enlarge the $r_1$ region.}
    \label{fig:distance}
\end{figure}

At low incident energy, the inclusive $P(r)$ deviates markedly from the smooth, continuous behavior expected for an uncorrelated Fermi gas.
Instead, it exhibits a highly structured profile characterized by three well-defined Gaussian components, as shown in Fig.~\ref{fig:distance}(a).
The solid line in Fig.~\ref{fig:distance}(a) represents a three-Gaussian fit, yielding peak positions at

\begin{equation}
    r_1 \approx 1.74\ \text{fm},\quad r_2 \approx 3.31\ \text{fm},\quad r_3 \approx 5.09\ \text{fm},
    \label{eq:peaks}
\end{equation}
Each fitted peak position coincides with a local maximum of the calculated $P(r)$.

The three peaks correspond to three distinct characteristic distances.
$r_1$ reflects the short-range inter-nucleon spacing.
Although its position retains a residual system dependence at low incident energy, in all four systems it grows with energy toward $\rho_0^{-1/3} \approx 1.8$~fm~\cite{cai_nucleon_2026}, reaching $1.83$--$1.86$~fm at the highest energies studied.
$r_2 = 3.31$~fm corresponds to an intermediate distance arising jointly from the nearest-neighbor $\alpha$--$\alpha$ structure and the finite-size constraint of the nucleus.
And $r_3 = 5.09$~fm reflects the long-range cluster--cluster correlation extending over distances comparable to the size of the system.

The intermediate-range peak at $r_2$ arises from the nearest-neighbor $\alpha$--$\alpha$ correlation and also carries a finite-size contribution.
Together with the long-range $\alpha$--$\alpha$ correlation at $r_3$, it defines the characteristic distances of the $\alpha$-cluster structure.
At low energy, the $r_2$ peak is dominated by $\alpha$-cluster structural correlations.
As the incident energy increases, this structural component is progressively suppressed.
Nevertheless, a maximum near $r_2$ persists because a purely finite-size contribution is also present at this distance.
Nucleons confined within a finite volume necessarily produce a two-body distance maximum at an intermediate distance set by the system size, regardless of any clustering.
A uniform distribution confined to a finite volume has a two-point distance distribution that vanishes at both zero separation and at twice the system radius, and must therefore possess an intermediate maximum.

The $r_2$ peak does not simply disappear across the transition.
What remains at this distance is a finite-size contribution that is present at all energies and carries no information about clustering.
Its persistence at high incident energy does not by itself establish that the $\alpha$--$\alpha$ correlations survive.
In the inclusive distribution, pairs at this distance also involve nucleons of the surrounding medium, and the question is settled only by the exclusive channel examined next.

The genuine microscopic hallmark of the nucleon liquid phase is correspondingly not this residual finite-size contribution but the growth of the $r_1$ peak.
The peak grows with increasing incident energy while its position moves toward the mean inter-nucleon distance at nuclear saturation density ($\rho_0^{-1/3} \approx 1.8$~fm~\cite{cai_nucleon_2026}), so that both its height and its position reflect short-range correlation within the nucleon medium rather than any clustering.
For the $^{28}$Si + $^{12}$C system, Fig.~\ref{fig:distance}(b) shows this growth directly.
The finite-size contribution arises solely from the spatial extent of the system.
Its existence can be established without specifying a value for the confinement radius.
We therefore make no quantitative use of such an estimate.

\subsection{Exclusive $\alpha$--$\alpha$ correlation function}

\begin{figure}[htbp]
    \centering
    \includegraphics[width=\columnwidth]{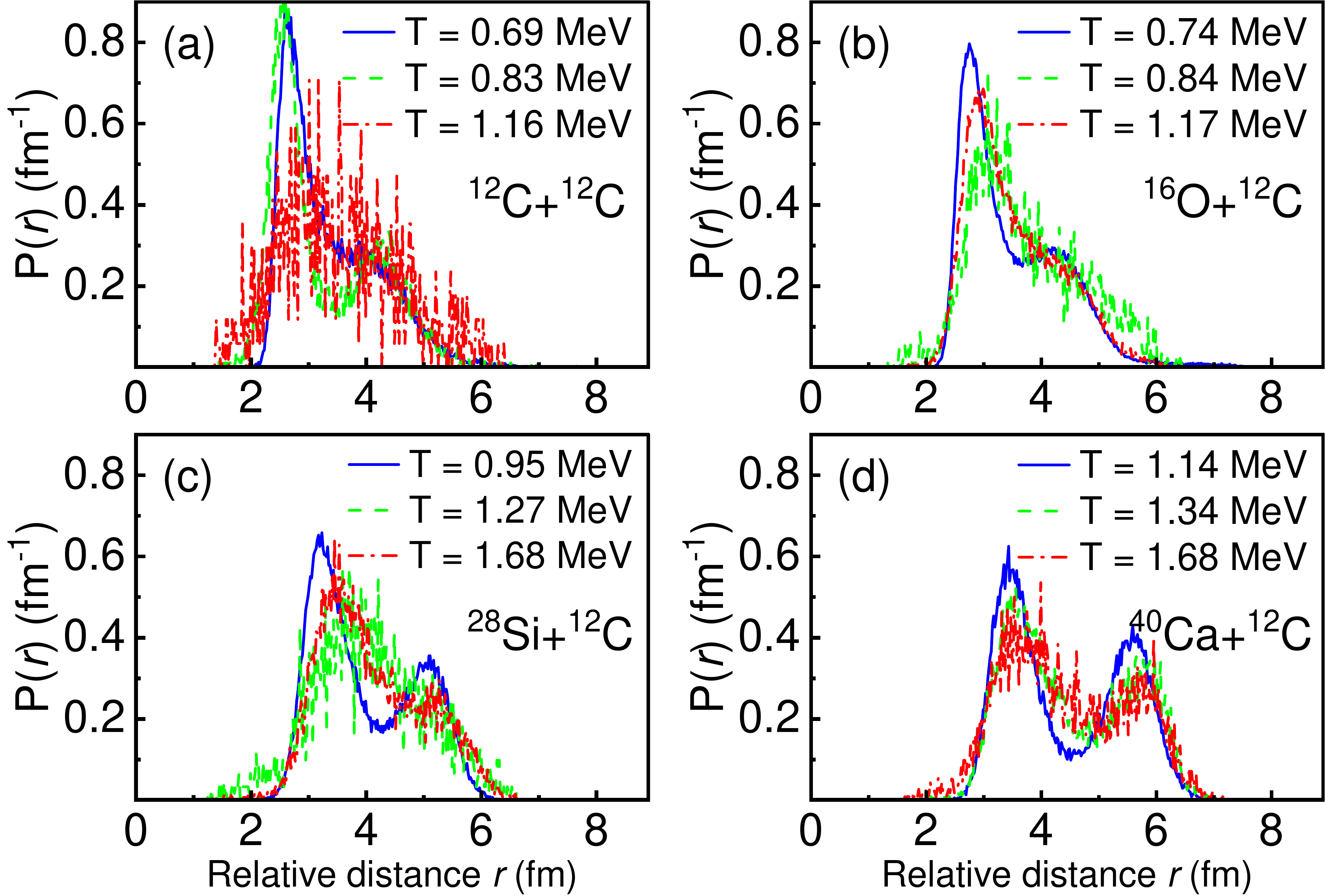}
    \caption{Energy evolution of the exclusive $\alpha$--$\alpha$ two-body spatial correlation function $P(r)$ for four $\alpha$-conjugate systems: (a) $^{12}\text{C}+^{12}\text{C}$, (b) $^{16}\text{O}+^{12}\text{C}$, (c) $^{28}\text{Si}+^{12}\text{C}$, and (d) $^{40}\text{Ca}+^{12}\text{C}$. The color scale (blue $\rightarrow$ green $\rightarrow$ red) denotes increasing incident energy, from the low-energy solid phase to the high-energy liquid phase.}
    \label{fig:exclusive_evolution}
\end{figure}

We therefore examine the exclusive $\alpha$--$\alpha$ correlation function $P(r)$ and its energy evolution in Fig.~\ref{fig:exclusive_evolution}.
To quantify this evolution, we characterize the exclusive $\alpha$--$\alpha$ distributions by their two leading peaks.
Two trends emerge as the incident energy increases from 2.0 to 7.0~MeV/u.
First, the abundance of the $\alpha$--$\alpha$ correlation collapses.
As quantified in Sec.~\ref{sec:quant}, the mean number of $\alpha$ clusters per event falls by 74\%--90\% in the four systems over this energy interval.
Because the number of $\alpha$--$\alpha$ pairs grows roughly as the square of the $\alpha$ multiplicity, it falls by more than an order of magnitude.
Nucleons are progressively transferred out of the clustered state as the system dissociates predominantly into the non-clustered nucleon medium.

Second, the characteristic distances of the remaining $\alpha$--$\alpha$ correlation change far less than its abundance.
Relative to the lowest energy available for each system, the nearest-neighbor peak position $\mu_1$ changes by less than 9\% depending on the system, while its width broadens by 35\%--53\%.
Table~\ref{Table:quant} lists the numerical values for all four systems.
The nearest-neighbor peak position is thus far more stable than either the abundance or the width of the correlation.
In contrast to the inclusive distribution, in which the long-range structure is progressively washed out with increasing energy, the nearest-neighbor $\alpha$--$\alpha$ correlation remains identifiable across the energy range studied.
The two channels therefore measure different aspects of the same system. The exclusive $\alpha$--$\alpha$ correlation tracks the spatial arrangement of the pairs of $\alpha$ clusters themselves, whereas the inclusive $P(r)$ becomes progressively dominated by the surrounding nucleon medium as the $\alpha$ multiplicity falls.

\subsection{Cluster abundance and characteristic distances}
\label{sec:quant}

\begin{figure}[htbp]
    \centering
    \includegraphics[width=\columnwidth]{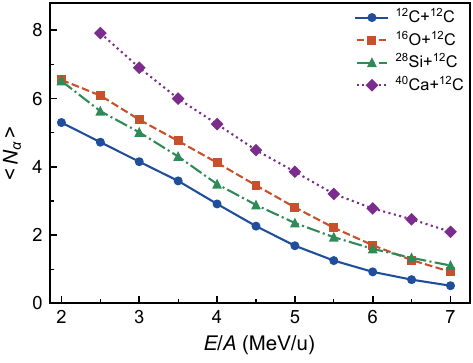}
    \caption{Energy evolution of the mean number of $\alpha$ clusters per event, $\langle N_\alpha\rangle$, for the four systems.
    As the incident energy increases from 2.0 to 7.0~MeV/u, $\langle N_\alpha\rangle$ falls by 74\%--90\%. For $^{40}$Ca+$^{12}$C the range starts at 2.5~MeV/u.}
    \label{fig:alpha_number}
\end{figure}

Table~\ref{Table:quant} lists the quantities extracted from the exclusive $\alpha$--$\alpha$ distribution, namely the $\alpha$ cluster abundance, the peak positions and the peak widths, while Fig.~\ref{fig:alpha_number} shows the energy evolution of the abundance.
The transition temperatures of the four systems, established in our previous study~\cite{SLPT2026}, are $T_c \approx 0.56$, 0.60, 1.20 and 1.29~MeV for $^{12}$C+$^{12}$C, $^{16}$O+$^{12}$C, $^{28}$Si+$^{12}$C and $^{40}$Ca+$^{12}$C, respectively.

The main quantitative result is the contrast between the abundance of the $\alpha$ clusters and the characteristic distances of that correlation.
The mean number of $\alpha$ clusters per event, $\langle N_\alpha\rangle$, falls from $5.30$, $6.55$, $6.51$ and $7.92$ at the lowest energy available for each system to $0.51$, $0.92$, $1.11$ and $2.09$ at 7~MeV/u for $^{12}$C+$^{12}$C, $^{16}$O+$^{12}$C, $^{28}$Si+$^{12}$C and $^{40}$Ca+$^{12}$C respectively, that is, a depletion of $90\%$, $86\%$, $83\%$ and $74\%$.
The nearest-neighbor position $\mu_1$ changes from 2.71, 2.81, 3.32 and 3.55~fm at the lowest energy of each system to 2.66, 3.03, 3.57 and 3.75~fm at 6~MeV/u, a change of less than 9\%. The position of the next-nearest-neighbor peak, $\mu_2$, stays within 3.82--4.14, 3.97--4.24, 4.94--5.15 and 5.49--5.65~fm over the same range (Table~\ref{Table:quant}).
The relative depletion in the number of $\alpha$ clusters is therefore much larger than the relative change of the nearest-neighbor distance.

The characteristic $\alpha$--$\alpha$ distance is a property of the system rather than a universal constant. Both distances grow with the size of the system. At 2.5~MeV/u, $\mu_1$ and $\mu_2$ increase from 2.68 and 3.85~fm in $^{12}$C+$^{12}$C to 3.55 and 5.49~fm in $^{40}$Ca+$^{12}$C (Table~\ref{Table:quant}).

\begin{table*}[t]
    \caption{Quantitative characterization of the exclusive $\alpha$--$\alpha$ correlation for the four systems.
    The three lighter systems are listed from 2.0 to 7.0~MeV/u and $^{40}$Ca+$^{12}$C from 2.5 to 7.0~MeV/u.
    $E$ is the incident energy per nucleon, $\langle N_\alpha\rangle$ the mean number of $\alpha$ clusters per event, $\mu_1$ and FWHM$_1$ the position and full width at half maximum of the nearest-neighbor peak, and $\mu_2$ the position of the next-nearest-neighbor peak in the two-Gaussian decomposition.
    Because the exclusive $\alpha$--$\alpha$ sample becomes sparse at the highest energies, $\mu_1$, FWHM$_1$ and $\mu_2$ are quoted only up to 6~MeV/u. The directly counted $\langle N_\alpha\rangle$ is tabulated over the full range. No uncertainties are quoted.}

    \begin{tabular*}{16cm} {@{\extracolsep{\fill}} lccccc}
    \toprule
    System & $E$ & $\langle N_\alpha\rangle$ & $\mu_1$ & FWHM$_1$ & $\mu_2$ \\
      & MeV/u &  & fm & fm & fm \\
    \midrule
    $^{12}$C+$^{12}$C ($^{24}$Mg) & 2.0 & 5.30 & 2.705 & 0.487 & 3.816 \\
       & 2.5 & 4.72 & 2.680 & 0.471 & 3.852 \\
       & 3.0 & 4.15 & 2.633 & 0.472 & 3.906 \\
       & 3.5 & 3.59 & 2.590 & 0.475 & 3.966 \\
       & 4.0 & 2.91 & 2.563 & 0.492 & 4.044 \\
       & 4.5 & 2.26 & 2.567 & 0.536 & 4.099 \\
       & 5.0 & 1.69 & 2.588 & 0.619 & 4.142 \\
       & 5.5 & 1.25 & 2.629 & 0.680 & 3.962 \\
       & 6.0 & 0.92 & 2.664 & 0.656 & 3.859 \\
       & 6.5 & 0.69 & --- & --- & --- \\
       & 7.0 & 0.51 & --- & --- & --- \\
    \midrule
    $^{16}$O+$^{12}$C ($^{28}$Si) & 2.0 & 6.55 & 2.813 & 0.540 & 4.014 \\
       & 2.5 & 6.09 & 2.823 & 0.536 & 4.020 \\
       & 3.0 & 5.38 & 2.823 & 0.522 & 4.026 \\
       & 3.5 & 4.76 & 2.827 & 0.550 & 4.007 \\
       & 4.0 & 4.12 & 2.840 & 0.594 & 3.971 \\
       & 4.5 & 3.46 & 2.875 & 0.653 & 3.976 \\
       & 5.0 & 2.81 & 2.925 & 0.707 & 3.985 \\
       & 5.5 & 2.21 & 2.990 & 0.776 & 4.113 \\
       & 6.0 & 1.70 & 3.029 & 0.827 & 4.239 \\
       & 6.5 & 1.27 & --- & --- & --- \\
       & 7.0 & 0.92 & --- & --- & --- \\
    \midrule
    $^{28}$Si+$^{12}$C ($^{40}$Ca) & 2.0 & 6.51 & 3.318 & 0.885 & 4.939 \\
       & 2.5 & 5.63 & 3.454 & 0.940 & 5.083 \\
       & 3.0 & 5.01 & 3.488 & 0.969 & 5.115 \\
       & 3.5 & 4.30 & 3.495 & 1.014 & 5.099 \\
       & 4.0 & 3.49 & 3.543 & 1.079 & 5.154 \\
       & 4.5 & 2.88 & 3.558 & 1.100 & 5.123 \\
       & 5.0 & 2.35 & 3.529 & 1.090 & 5.001 \\
       & 5.5 & 1.94 & 3.613 & 1.256 & 5.121 \\
       & 6.0 & 1.59 & 3.567 & 1.201 & 4.952 \\
       & 6.5 & 1.33 & --- & --- & --- \\
       & 7.0 & 1.11 & --- & --- & --- \\
    \midrule
    $^{40}$Ca+$^{12}$C ($^{52}$Fe) & 2.5 & 7.92 & 3.545 & 0.921 & 5.493 \\
       & 3.0 & 6.90 & 3.607 & 0.970 & 5.551 \\
       & 3.5 & 6.00 & 3.663 & 1.032 & 5.596 \\
       & 4.0 & 5.25 & 3.693 & 1.092 & 5.631 \\
       & 4.5 & 4.49 & 3.697 & 1.138 & 5.635 \\
       & 5.0 & 3.85 & 3.716 & 1.201 & 5.640 \\
       & 5.5 & 3.20 & 3.737 & 1.230 & 5.647 \\
       & 6.0 & 2.78 & 3.749 & 1.277 & 5.645 \\
       & 6.5 & 2.46 & --- & --- & --- \\
       & 7.0 & 2.09 & --- & --- & --- \\
    \bottomrule
    \end{tabular*}
    \label{Table:quant}
\end{table*}

\section{Discussion}

The cluster analysis directly measures a reduction in the number of $\alpha$ clusters and a corresponding transfer of nucleons into the non-clustered nucleon medium.
It does not by itself establish whether individual clusters break apart or whether nucleons are lost from the system, but in the present calculation the second possibility is not open.
The total Hamiltonian is conserved and no emission or decay channel is included (Sec.~\ref{sec:noemission}).
The observed collapse in the number of $\alpha$ clusters therefore documents a redistribution of nucleons inside the system.

Our previous study~\cite{SLPT2026} read the progressive washing out of the long-range peaks as the spatial fingerprint of a melting transition.
However, the signals reported there, including the specific heat and caloric curve anomalies, the growth of the number of fragment species and of the information entropy, and the disappearance of the long-range peaks, do not distinguish between two distinct mechanisms.

The long-range correlation may be lost either because the clusters themselves dissociate, or because their positions within the ordered arrangement become randomized.
The exclusive $\alpha$--$\alpha$ channel, built from the $\alpha$-cluster centroids alone, separates the two, and the present results assign the dominant role to the former.
Against a collapse of 74\%--90\% in the number of $\alpha$ clusters, the nearest-neighbor $\alpha$--$\alpha$ distance changes by less than 9\%.
The distance distribution broadens by 35\%--53\%.
The broadening is a spread in the nearest-neighbor distances of the surviving pairs rather than a displacement of their mean spacing, since a randomization of the positions within the arrangement, the alternative considered above, would move or wash out the peak itself, whereas $\mu_1$ changes by less than 9\%.
What changes decisively is the number of clusters in the ordered arrangement, not its characteristic distance.
The inclusive analysis of Ref.~\cite{SLPT2026} thus shows that the long-range structure is lost, while the exclusive analysis identifies what is lost with it.
The depletion develops continuously with incident energy, which is consistent with the enhanced thermal motion and collision-induced fluctuations progressively driving nucleons out of their cluster-localized configurations.
A gradual rather than abrupt solid-liquid transition is a property of finite systems, in which the discontinuity of the bulk transition is smoothed out~\cite{Cheng1992}.

\section{Summary and conclusions}

In this work, we have investigated the structural evolution of $\alpha$-cluster matter in four $\alpha$-conjugate nuclear collision systems using a microscopic transport model together with an exclusive two-body spatial correlation analysis that isolates the $\alpha$ clusters.
Our analysis extends the two-body correlation function $P(r)$ of our previous study~\cite{SLPT2026}, whose inclusive variant sums over all constituents, so that the $\alpha$--$\alpha$ pairs cannot be separated from the pairs of all other kinds.
The exclusive $\alpha$-$\alpha$ variant used here retains only the pairs of $\alpha$ clusters and exposes the correlation among them.

Our main results concern the characteristic distances of the $\alpha$-$\alpha$ correlation.
All four systems exhibit nearest-neighbor $\alpha$-$\alpha$ correlation and next-nearest-neighbor inter-cluster structure.
The nearest-neighbor distance increases from about 2.7~fm in the lightest system studied to about 3.5~fm in the heaviest, and the next-nearest-neighbor distance from about 3.8 to about 5.6~fm.
These distances are therefore not universal constants but properties that vary with the size of the system.

Two properties of the $\alpha$--$\alpha$ correlation, its abundance and its characteristic distances, evolve very differently with incident energy.
As the incident energy increases from 2.0 to 7.0~MeV/u, the mean number of $\alpha$ clusters per event falls by 74\%--90\% in all four systems, whereas both the nearest-neighbor and next-nearest-neighbor distances change only by less than 9\% (Table~\ref{Table:quant}).
The $\alpha$--$\alpha$ correlation therefore persists even as the number of clusters declines.
This persistence does not conflict with the fading of the long-range structure in the inclusive distribution, which describes the system as a whole, whereas the exclusive correlation isolates the surviving pairs of $\alpha$ clusters.

A careful attribution of the three peaks of the inclusive $P(r)$ shows that they do not share a single physical origin.
The short-range peak $r_1$ converges with increasing incident energy to the mean inter-nucleon distance at nuclear saturation density ($\rho_0^{-1/3} \approx 1.8$~fm~\cite{cai_nucleon_2026}) in all four systems, and is a signature of the nucleon medium rather than of clustering.
The intermediate-distance feature near 3~fm is dominated by $\alpha$--$\alpha$ nearest-neighbor structure in the low-energy solid phase.
The $\alpha$-$\alpha$ contribution to this feature is progressively suppressed as the incident energy increases, while the finite-size contribution, set by the spatial extent of the system, remains unchanged.
At the highest energies this intermediate-distance feature is therefore dominated by the finite-size contribution, although a residual $\alpha$--$\alpha$ component is still present.
That contribution carries no information about clustering and cannot by itself be taken as evidence that the $\alpha$--$\alpha$ correlations survive.
The nucleons transferred out of the clustered state enter this same medium, whose short-range signature is the peak $r_1$.

At densities close to saturation, without appreciable expansion of the system, the number of $\alpha$ clusters falls steeply as the incident energy increases, while the distances that characterize the surviving $\alpha$--$\alpha$ correlation change only slightly.
This contrast identifies the dominant mechanism of the transition as the dissolution of the individual clusters rather than as a randomization of their positions within the ordered arrangement.

\begin{acknowledgments}
This work was supported by the National Natural Science Foundation of China (Grants No. 12475134, No. 12441505, No. 12147101, No. 11975210, and No. U1832129), the Strategic Priority Research Program of the Chinese Academy of Sciences (Grant No. XDB34030000), the Youth Innovation Promotion Association CAS (Grant No. 2017309), and the National Key Research and Development Program of China (Grant No. 2022YFA1602404).
\end{acknowledgments}

\bibliography{refer}

\end{document}